\documentclass[12pt]{article}
\usepackage[dvips]{graphicx}
\usepackage{amsmath,amssymb,amsthm,bm}
\bmdefine{\BX}{{\bm X}}
\bmdefine{\Bx}{{\bm x}}
\bmdefine{\By}{{\bm y}}
\bmdefine{\BY}{{\bm Y}}
\bmdefine{\Bz}{{\bm z}}
\bmdefine{\Ba}{{\bm a}}
\bmdefine{\Bd}{{\bm d}}
\bmdefine{\Bu}{{\bm u}}
\bmdefine{\Bh}{{\bm h}}
\bmdefine{\Bm}{{\bm m}}
\bmdefine{\Bi}{{\bm i}}
\bmdefine{\Bc}{{\bm c}}
\bmdefine{\BT}{{\bm T}}
\bmdefine{\Bp}{{\bm p}}
\bmdefine{\Bs}{{\bm s}}
\bmdefine{\Btheta}{{\bm \theta}}
\bmdefine{\Bmu}{{\bm \mu}}
\bmdefine{\Bt}{{\bm t}}
\bmdefine{\Blambda}{{\bm \lambda}}
\bmdefine{\Bbeta}{{\bm \beta}}
\bmdefine{\Bpsi}{{\bm \psi}}
\bmdefine{\Bpi}{{\bm \pi}}
\bmdefine{\Bv}{{\bm v}}
\bmdefine{\Bzero}{{\bm 0}}
\bmdefine{\Bone}{{\bm 1}}
\bmdefine{\Bvarepsilon}{{\bm \varepsilon}}

\bmdefine{\BI}{{\rm {\bm I}}}
\bmdefine{\BV}{{\rm {\bm V}}}

\theoremstyle{definition}
\newtheorem{theorem}{Theorem}[section]

\newtheorem{proposition}[theorem]{Proposition}

\newtheorem{definition}[theorem]{Definition}
\newtheorem{remark}[theorem]{Remark}

\title{
Use of indicator functions to enumerate cross-array designs without
direct product structure
}
\author{Satoshi Aoki%
\thanks{Graduate School of Science, Faculty of Science, Kobe
University.}
and 
Masayuki Noro%
\thanks{Department of Mathematics, College of Science, Rikkyo University}
}
\date{}

\begin{document}
\maketitle

\begin{abstract}
\noindent
Use of polynomial indicator functions to enumerate fractional factorial
 designs with given properties is first introduced by Fontana,
 Pistone and Rogantin (2000) for two-level factors, and generalized by
 Aoki (2019) for multi-level factors. In this paper, we apply this
 theory to enumerate cross-array designs. 
For the experiments of several control factors and noise factors, 
use of the cross-array designs with direct product structure is
 widespread as an effective robust strategy in Taguchi method. In this
 paper, we relax this direct product structure to reduce the size of the
 designs. We obtain $24$-runs cross-array designs without direct product
 structure with some desirable properties for $6$ control factors and
 $3$ noise factors, each with
 two-levels, instead of $32$-runs design that is widely used.  

\noindent
Keywords:\ 
Computational algebraic statistics, 
Cross-array designs,
Fractional factorial designs, 
Gr\"obner bases, 
Indicator functions, 
Taguchi methods.
\end{abstract}

\section{Introduction}
\label{sec:intro}
Suppose we plan to conduct designs of experiment for 
several control factors and several noise factors. In the theory of
quality engineering, which is one of the important fields in designs of
experiments, use of {\it cross-array designs} is recommended as an effective
robust strategy.
For example, suppose we have $6$ control factors $x_1,\ldots,x_6$ and
$3$ noise factors $y_1,\ldots,y_3$, each with $2$ levels $\{-1,1\}$.
In this case, we can allocate the control factors to the
orthogonal array $L_8$ and the noise factors to the orthogonal array
$L_4$, and obtain a cross-array design shown in Table
\ref{tbl:6-3-cross-array}. 
\begin{table}[htbp]
\begin{center}
\caption{Cross-array design for $6$ control factors and $3$ noise
 factors.}
\label{tbl:6-3-cross-array}
\begin{tabular}{rrrrrrc|c|c|c|c|}\cline{8-11}
\multicolumn{6}{c}{} & $y_1$ & 
\multicolumn{1}{|r|}{$-1$}
 & \multicolumn{1}{|r|}{$-1$}
 & \multicolumn{1}{|r|}{$1$}
 & \multicolumn{1}{|r|}{\ \ $1$\ }\\
\multicolumn{6}{c}{} & $y_2$ & 
\multicolumn{1}{|r|}{$-1$}
 & \multicolumn{1}{|r|}{$1$}
 & \multicolumn{1}{|r|}{$-1$}
 & \multicolumn{1}{|r|}{$1$}\\
$x_1$ & $x_2$ & $x_3$ & $x_4$ & $x_5$ & $x_6$ & $y_3$ & 
\multicolumn{1}{|r|}{$-1$}
 & \multicolumn{1}{|r|}{$1$}
 & \multicolumn{1}{|r|}{$-1$}
 & \multicolumn{1}{|r|}{$1$}\\ \hline 
$-1$ & $-1$ & $-1$ & $1$ & $1$ & $1$ & & $\circ$ & $\circ$ & $\circ$ & $\circ$\\ \hline
$-1$ & $-1$ & $1$ & $1$ & $-1$ & $-1$ & & $\circ$ & $\circ$ & $\circ$ & $\circ$\\ \hline
$-1$ & $1$ & $-1$ & $-1$ & $1$ & $-1$ & & $\circ$ & $\circ$ & $\circ$ & $\circ$\\ \hline
$-1$ & $1$ & $1$ & $-1$ & $-1$ & $1$ & & $\circ$ & $\circ$ & $\circ$ & $\circ$\\ \hline
$1$ & $-1$ & $-1$ & $-1$ & $-1$ & $1$ & & $\circ$ & $\circ$ & $\circ$ & $\circ$\\ \hline
$1$ & $-1$ & $1$ & $-1$ & $1$ & $-1$ & & $\circ$ & $\circ$ & $\circ$ & $\circ$\\ \hline
$1$ & $1$ & $-1$ & $1$ & $-1$ & $-1$ & & $\circ$ & $\circ$ & $\circ$ & $\circ$\\ \hline
$1$ & $1$ & $1$ & $1$ & $1$ & $1$ & & $\circ$ & $\circ$ & $\circ$ & $\circ$\\ \hline
\end{tabular}
\end{center}
\end{table}
In Table \ref{tbl:6-3-cross-array}, we denote each observation by the 
symbol ``$\circ$''.
Use of the cross-array designs is a popular strategy in the context of
robust parameter designs in Taguchi methods. See Chapter 15 of
\cite{Taguchi-2005} or Chapter 11 of \cite{Wu-Hamada-2009}. 
We can also find various orthogonal arrays such as $L_8$
and $L_4$ in these textbooks. In this field, the cross-array design is
also called an inner-outer array design (or an inner-outer layout), 
where the array for the control
factors ($L_8$ in Table \ref{tbl:6-3-cross-array}) is called an inner
array, and the array for the noise factors  ($L_4$ in Table
\ref{tbl:6-3-cross-array}) is called an outer array. From the
observations obtained in the cross-array design, we can estimate all main
effects of the control factors and the noise factors, and all
interaction effects between the control factors and the noise
factors. This is the principle merit to use the cross-array designs. See
textbooks such as \cite{Taguchi-2005} or \cite{Wu-Hamada-2009} for
detail. 

Obviously the cross-array designs have a direct product structure
between the inner array and the outer array. In this paper, we relax
this direct product structure to reduce the size of the designs. 
Note that the reduction in size of the designs is always of interest for
cost saving.  
To understand the idea in this paper clearly, we rewrite Table 
\ref{tbl:6-3-cross-array} as Table 
\ref{tbl:6-3-cross-array-2} and emphasize the direct product structure. 
\begin{table}[htbp]
\begin{center}
\caption{Same cross-array design of Table \ref{tbl:6-3-cross-array}.}
\label{tbl:6-3-cross-array-2} 
\begin{tabular}{rrrrrr|r|r|r|r|r|r|r|r|}\cline{6-14}
\multicolumn{5}{c}{} & \multicolumn{1}{|c|}{$y_1$} & $-1$ & $-1$ & $-1$ & $-1$ & $1$ & $1$ &
 $1$ & $1$\\
\multicolumn{5}{c}{} & \multicolumn{1}{|c|}{$y_2$} & $-1$ & $-1$ & $1$ & $1$ & $-1$ & $-1$ &
 $1$ & $1$\\ 
\multicolumn{5}{c}{} & \multicolumn{1}{|c|}{$y_3$} & $-1$ & $1$ & $-1$ &
 $1$ & $-1$ & $1$ & $-1$ & $1$\\ \cline{6-6}
$x_1$ & $x_2$ & $x_3$ & $x_4$ & $x_5$ & $x_6$ & & & & & & & & \\ \hline
$-1$ & $-1$ & $-1$ & $1$ & $1$ & $1$ & & $\circ$ & $\circ$ & & $\circ$
 & & & $\circ$\\ \hline
$-1$ & $-1$ & $1$ & $1$ & $-1$ & $-1$ & & $\circ$ & $\circ$ & & $\circ$
 & & & $\circ$\\ \hline
$-1$ & $1$ & $-1$ & $-1$ & $1$ & $-1$ & & $\circ$ & $\circ$ & & $\circ$
 & & & $\circ$\\ \hline
$-1$ & $1$ & $1$ & $-1$ & $-1$ & $1$ & & $\circ$ & $\circ$ & & $\circ$
 & & & $\circ$\\ \hline
$1$ & $-1$ & $-1$ & $-1$ & $-1$ & $1$ & & $\circ$ & $\circ$ & & $\circ$
 & & & $\circ$\\ \hline
$1$ & $-1$ & $1$ & $-1$ & $1$ & $-1$ & & $\circ$ & $\circ$ & & $\circ$
 & & & $\circ$\\ \hline
$1$ & $1$ & $-1$ & $1$ & $-1$ & $-1$ & & $\circ$ & $\circ$ & & $\circ$
 & & & $\circ$\\ \hline
$1$ & $1$ & $1$ & $1$ & $1$ & $1$ & & $\circ$ & $\circ$ & & $\circ$
 & & & $\circ$\\ \hline
\end{tabular}
\end{center}
\end{table}
In this study, 
instead of this $32$-runs design, 
we try to construct $24$-runs designs with some desirable properties. 
As we see in Section \ref{sec:24-runs-fractions}, we construct $24$-runs
designs such as Table \ref{tbl:6-3-new-design} using tools of algebraic
statistics. In fact, this (and {\it only this}) $24$-runs design has
some desirable properties we consider, which we show in Section 3. 
\begin{table}[htbp]
\begin{center}
\caption{$24$-runs cross-array designs for $6$ control factors and $3$
 noise factors.}
\label{tbl:6-3-new-design} 
\begin{tabular}{rrrrrr|r|r|r|r|r|r|r|r|}\cline{6-14}
\multicolumn{5}{c}{} & \multicolumn{1}{|c|}{$y_1$} & $-1$ & $-1$ & $-1$ & $-1$ & $1$ & $1$ &
 $1$ & $1$\\
\multicolumn{5}{c}{} & \multicolumn{1}{|c|}{$y_2$} & $-1$ & $-1$ & $1$ & $1$ & $-1$ & $-1$ &
 $1$ & $1$\\ 
\multicolumn{5}{c}{} & \multicolumn{1}{|c|}{$y_3$} & $-1$ & $1$ & $-1$ &
 $1$ & $-1$ & $1$ & $-1$ & $1$\\ \cline{6-6}
$x_1$ & $x_2$ & $x_3$ & $x_4$ & $x_5$ & $x_6$ & & & & & & & & \\ \hline
$-1$ & $-1$ & $-1$ & $1$ & $1$ & $1$  & $\circ$ & $\circ$ & & & & & & $\circ$\\ \hline
$-1$ & $-1$ & $1$ & $1$ & $-1$ & $-1$ & & & $\circ$ & & & $\circ$ & $\circ$ & \\ \hline
$-1$ & $1$ & $-1$ & $-1$ & $1$ & $-1$ & & & $\circ$ & & $\circ$ & & & $\circ$ \\ \hline
$-1$ & $1$ & $1$ & $-1$ & $-1$ & $1$  & & $\circ$ & & $\circ$ & $\circ$ & & & \\ \hline
$1$ & $-1$ & $-1$ & $-1$ & $-1$ & $1$ & & & & $\circ$ & $\circ$ & & $\circ$ & \\ \hline
$1$ & $-1$ & $1$ & $-1$ & $1$ & $-1$  & $\circ$ & & & $\circ$ & & $\circ$ & & \\ \hline
$1$ & $1$ & $-1$ & $1$ & $-1$ & $-1$  & & $\circ$ & $\circ$ & & & $\circ$ & & \\ \hline
$1$ & $1$ & $1$ & $1$ & $1$ & $1$     & $\circ$ & & & & & & $\circ$ & $\circ$\\ \hline
\end{tabular}
\end{center}
\end{table}

To construct designs with some given properties, we make use of 
{\it polynomial indicator functions of designs}. This algebraic technique is
first introduced in \cite{Fontana-Pistone-Rogantin-2000} for designs of 
two-level factors, and is generalized to multi-level cases in
\cite{Aoki-2019}. By the generalization by \cite{Aoki-2019}, we can
construct designs
with given
properties for any number of factors with any set of levels {\it in theory}.  
However, in practice, the computational feasibility becomes a
problem. For example, we enumerate all orthogonal fractions of
$2^4\times 3$ designs with strength $3$ in \cite{Aoki-2019}, but fail to
enumerate all orthogonal fractions of $2^4\times 3$ designs with
strength $2$. (We give a definition of a strength of orthogonality in Section 
\ref{sec:24-runs-fractions}.)
The latter problem is solved in \cite{Aoki-Noro-2022} using
technique of a primary decomposition of polynomial ideals. 
These problems correspond to solving algebraic equations for $2^4\times
3 = 48$ variables with some constraints (\cite{Aoki-2019} and
\cite{Aoki-Noro-2022}).
In the same way, we can define
the polynomial ideal for the problems of $6$ control factors and $3$
noise factors as an ideal with $2^9=512$ variables. Obviously, such a
naive approach fail to solve 
in realistic time. As we see in Section 3, we
define the ideal with $2^6=64$ variables for this problem, which we can
solve by an usual laptop PC. 

The content of this paper is as follows. In Section 2, we summarize the
theory of the indicator function and its use to construct designs. 
Because we apply this theory to the cross-array designs, we
summarize the general theory given by \cite{Aoki-2019} in the cross-array
settings. In Section 3, we consider the problem for $6$ control factors
and $3$ noise factors. We show the properties of $24$-runs designs such
as Table \ref{tbl:6-3-new-design} and show how to construct them. 

\section{Constructing fractions using indicator functions}
\label{sec:indicator-function}
In this section, we summarize the theory on the indicator function of
designs and its use to construct designs. 
The arguments are based on the theory of the interpolatory
polynomials on designs, which is one of the first applications of
Gr\"obner bases to statistics by \cite{Pistone-Wynn-1996}. 
The arguments in this section are given in \cite{Aoki-2019} in general
settings. We give the arguments in the setting of the cross-array designs in
this section.

Let $x_1,.\ldots,x_p$ be $p$ control factors. Let $A_j \subset
\mathbb{Q}$ be a level set of a control factor $x_j$ for $j=1,\ldots,p$,
where $\mathbb{Q}$ denotes the field of rational numbers. A full
factorial design of the control factors $x_1,\ldots,x_p$ is $D_x =
A_1 \times \cdots \times A_p$. A subset of $D_x$, $F_x \subset D_x$, is
called a fractional factorial design of the control factors. 
We also define similar materials for $q$ noise factors $y_1,\ldots,y_q$,
i.e., let $B_j \subset \mathbb{Q}$ be a level set of a noise factor
$y_j$ for $j = 1,\ldots,q$, $D_y = B_1\times\cdots\times B_q$ be a full
factorial design, and $F_y \subset D_y$ be a fractional factorial design
for the noise factors $y_1,\ldots,y_q$.

For fractional factorial designs $F_x\subset D_x$ and $F_y \subset D_y$, 
$F_x \times F_y$ is a cross-array design with a direct product
structure. 
In the theory of Taguchi method, use of direct product type 
cross-array designs $F_x\times F_y$
is recommended for 
{\it regular} fractional factorial designs $F_x$ and $F_y$.
In this paper, we consider
general fractional factorial designs of $F_x \times D_y$ for a regular
fractional factorial design $F_x \subset D_x$ as follows. 
We assume that the levels of the $r$\ $(r < p)$ 
control factors $x_{p-r+1},\ldots,x_p$ are determined from the
levels of the remaining $p-r$\ $(= s)$ control factors $x_1,\ldots,x_s$ by
\begin{equation}
 x_{s+j} = g_j(x_1,\ldots,x_s),\ j = 1,\ldots,r,
\label{eqn:def-relation}
\end{equation}
where $g_j \in \mathbb{Q}[x_1,\ldots,x_s]$ for $j = 1,\ldots,r$. 
Under this relation, the fractional factorial design
\[
 \{(x_1,\ldots,x_s,g_1(x_1,\ldots,x_s),\ldots,g_r(x_1,\ldots,x_s))\ |\
 (x_1,\ldots,x_s) \in A_1\times\cdots\times A_s\} \subset D_x
\] 
is called a regular fractional factorial design of $D_x$ with the
defining relation $g_1,\ldots,g_r$. 
\begin{remark}
In the textbooks such as \cite{Wu-Hamada-2009}
 or \cite{Taguchi-2005}, the regular fractional factorial design is 
 mainly explained for two or three level factors. For two-level cases,
 which we also consider in this paper, it is common to define such
 as $x_4 = g_1(x_1,x_2,x_3) = x_1x_2$ for the level $\{-1,1\}$. For
 three-level cases, it is also common to define by ``mod $3$ operation''
 such as 
\[
 x_4 = x_1 + x_2 + x_3\ (\mbox{mod}\ 3)
\]
for the level set $\{0,1,2\}$. 
Though our definition of the regular designs by (\ref{eqn:def-relation})
 seems different from the conventional definition given in the textbooks
 in this field, our definition is more general and holds for factors
 with different number of levels. Note that it is also possible to
 represent the above $x_4$ by the polynomial in $\mathbb{Q}[x_1,x_2,x_3]$ as
 the interpolation function on $\{0,1,2\}^3$ with the response
 $x_4$. 
\end{remark}

We write the full factorial design of $x_1,\ldots,x_s$ as 
$D_x^* = A_1\times \cdots \times A_s$ and consider fractional factorial
designs of $D_x^*\times D_y$ in this paper. Now we introduce an
indicator function on $D_x^*\times D_y$. 
\begin{definition}[\cite{Fontana-Pistone-Rogantin-2000}]
The indicator function of $F \subset D_x^*\times D_y$ 
is a response function $f$ on $D_x^*\times D_y$ satisfying
\[
 f(x_1,\ldots,x_s,y_1,\ldots,y_q) = \left\{
\begin{array}{ll}
1, & \mbox{if}\ (x_1,\ldots,x_s,y_1,\ldots,y_q) \in F,\\
0, & \mbox{if}\ (x_1,\ldots,x_s,y_1,\ldots,y_q) \in (D_x^*\times
 D_y)\setminus F.
\end{array}
\right.
\]
\label{def:indicator-function}
\end{definition}

To show the existence, uniqueness and construction of the polynomial indicator
function below, 
we prepare the set $L \subset
\mathbb{Z}^{s+q}_{\geq 0}$ by 
\[
 L = \left\{(a_1,\ldots,a_s,b_1,\ldots,b_q)\ \left|\ 
\begin{array}{l}
 0 \leq a_i \leq |A_i|-1,\ i = 1,\ldots,s,\\
 0 \leq b_j \leq |B_j|-1,\ j = 1,\ldots,q
\end{array}
\right.\right\},
\]
where $\mathbb{Z}_{\geq 0}$ be the set of nonnegative integers, and
$|A_i|$ and $|B_j|$ be the cardinality of $A_i$ and $B_j$, respectively.
By ordering the points of $D_x^*\times D_y$, we write
\[
 D_x^*\times D_y = \{\Bd_1,\ldots,\Bd_m\},
\]
where $m = \left(\prod_{j = 1}^s|A_j|\right)\left(\prod_{j =
1}^q|B_j|\right)$ is the size of $D_x^*\times D_y$, and 
\[
 \Bd_i = (x_{i1},\ldots,x_{is},y_{i1},\ldots,y_{iq}),\ i = 1,\ldots,m,
\]
where $x_{ij}$ and $y_{ij}$ are the levels of the factors $x_j$ and $y_j$
in the $i$th design point (i.e., $i$th run), respectively. 
For each $\Bc = (a_1,\ldots,a_s,b_1,\ldots,b_q) \in L$, write the
monomial as
\[
 \Bd_i^{\Bc} = x_{i1}^{a_1}\cdots x_{is}^{a_s}y_{i1}^{b_1}\cdots y_{iq}^{b_q}
\]
for $i = 1,\ldots,m$. By ordering the elements of $L$, we define a model
matrix of $D_x^*\times D_y$ by
\[
 X = \left[\Bd_i^{\Bc}\right]_{i=1,\ldots,m;\ \Bc\in L}.
\]
Note that $X$ is called a design matrix in Definition 26 of
\cite{Pistone-Riccomagno-Wynn-2001}, and is an $m\times m$ nonsingular
matrix (Theorem 26 of \cite{Pistone-Riccomagno-Wynn-2001}).
For the variables $\Bz = (x_1,\ldots,x_s,y_1,\ldots,y_q)$ and $\Bc =
(a_1,\ldots,a_s,b_1,\ldots,b_q) \in L$, we also write the monomial as
\[
 \Bz^{\Bc} = x_1^{a_1}\cdots x_s^{a_s}y_1^{b_1}\cdots y_q^{b_q}, 
\]
and write the indicator function of $F\subset D_x^*\times D_y$ as
\[
 f(x_1,\ldots,x_s,b_1,\ldots,b_q) = f(\Bz) = \sum_{\Bc \in
 L}\theta_{\Bc}\Bz^{\Bc}.
\]
Then the $m\times 1$ column vector $\Btheta =
\left[\theta_{\Bc}\right]_{\Bc \in L}$ is given by $\Btheta =
X^{-1}\Bu$, where $\Bu = \left[u_1,\ldots,u_m\right]^T \in \{0,1\}^m$ is
an $m\times 1$ column vector of a response on $D_x^*\times D_y =
\{\Bd_1,\ldots,\Bd_m\}$
satisfying
\begin{equation}
 u_i = \left\{\begin{array}{ll}
1, & \mbox{if}\ \Bd_i \in F,\\
0, & \mbox{if}\ \Bd_i \in (D_x^*\times D_y)\setminus F.
\end{array}
\right.
\label{eqn:def-u}
\end{equation}
See Theorem 26 of \cite{Pistone-Riccomagno-Wynn-2001} for detail.
An important fact is that the set $\{\Bz^{\Bc}\ |\ \Bc \in L\}$ is a
basis of $\mathbb{Q}[\Bz]/I(D_x^*\times D_y)$ as a $\mathbb{Q}$-vector
space, where $I(D_x^*\times D_y)$ is the design ideal of $D_x^*\times
D_y$, i.e., the 
set of polynomials in
$\mathbb{Q}[\Bz]$ which are $0$ at every point of $D_x^*\times
D_y$. This fact guarantees the existence and the uniqueness of the
polynomial indicator function of $F\subset D_x^*\times D_y$. See Theorem
15 of \cite{Pistone-Riccomagno-Wynn-2001} for detail. 

The fractional factorial designs with given properties can be
characterized as the zero points of the $0$-dimensional ideal defined
for the coefficients of the corresponding indicator function. Let ${\cal
G}$ be the reduced Gr\"obner basis of $I(D_x^*\times D_y)$ for some
monomial order on $\Bz$. Then a polynomial $f(\Bz) = \sum_{\Bc \in
L}\theta_{\Bc}\Bz^{\Bc}$ is an indicator function of some fractional
factorial design of $D_x^{*}\times D_y$ if and only if the system of
the algebraic equations
\begin{equation}
 \theta_{\Bc} = \mu_{\Bc},\ \ \Bc \in L
\label{eqn:sys-theta=mu}
\end{equation}
hold, where $\sum_{\Bc \in L}\mu_{\Bc}\Bz^{\Bc}$ is the standard form
of $\sum_{\Bc_1 \in L}\sum_{\Bc_2 \in
L}\theta_{\Bc_1}\theta_{\Bc_2}\Bz^{\Bc_1 + \Bc_2}$ with respect to
${\cal G}$ (Proposition 3.1 of \cite{Aoki-2019}). 
Therefore, if we give some properties for the fractional factorial
designs as $k$ constraints 
in the form of $C\Bu = \Bh$ for the corresponding response
vector $\Bu$ of (\ref{eqn:def-u}), where $C$ is a $k\times m$ matrix and
$\Bh$ is a $k\times 1$ column vector, 
we can define a polynomial ideal of $\mathbb{Q}[\Btheta]$ by
\begin{equation}
I = \left<\{\theta_{\Bc} - \mu_{\Bc},\ \Bc \in L\}, CX\Btheta - \Bh
\right>, 
\end{equation}
and the variety defined by the polynomial ideal $I$,
\[
 V(I) = \{\Btheta \in \mathbb{Q}^m\ |\ f(\Btheta) = 0,\ \forall f \in I\},
\]
corresponds to the set of the coefficients of the indicator functions 
of the fractional factorial designs with the given properties.
Note that $V(I)$ is finite, i.e., $I$ is a $0$-dimensional ideal. See
\cite{Aoki-2019} and 
\cite{Aoki-Noro-2022} for detail.

\section{$24$-runs cross-array designs for $6$ control factors and $3$ noise
 factors with two-levels}
\label{sec:24-runs-fractions}
In this section, we use the theory given in the previous section to the
problem of $24$-runs designs for $6$ control factors and $3$ noise
factors. 
Therefore the problem we consider corresponds to the case 
\[
 p=6,\ q=3,\ s=3,\ A_1=\cdots = A_6 = B_1 = \cdots = B_3 = \{-1,1\},\ m
 = 2^3\cdot 2^3 = 64.
\]
The defining relation (\ref{eqn:def-relation}) is given by
\[
x_4 = g_1(x_1,x_2,x_3) = x_1x_2,\ 
x_5 = g_2(x_1,x_2,x_3) = x_1x_3,\ 
x_6 = g_3(x_1,x_2,x_3) = x_2x_3. 
\]
Because we consider two-level cases, the algebraic
equations (\ref{eqn:sys-theta=mu}) reduce to
\begin{equation}
 \theta_{\Bc} = \sum_{\Bc' \in L}\theta_{\Bc'}\theta_{\Bc\triangle
 \Bc'}\ \ \Bc \in L,
\label{eqn:2-level-equations}
\end{equation}
where we denote by $\triangle$ the symmetric difference for $I,I' \in
L$, i.e., $I\triangle I' = (I\cup I')\setminus (I\cap I')$. See
Proposition 3.7 of \cite{Fontana-Pistone-Rogantin-2000} for detail. 

To consider the constraints on $\Bu$, we introduce notations of
{\it contingency tables} as follows. 
For each response $\Bu \in \{0,1\}^{64}$ on $D_x^{*}\times D_y =
\{-1,1\}^3\times \{-1,1\}^3$, we treat $\Bu$ as 
a $2^6$
contingency table with the set of cells ${\cal I} = \{1,2\}^6$ and write 
$\Bu = (u(\Bi))_{\Bi \in {\cal I}}$. We specify each cell $\Bi \in {\cal
I}$ by $\Bi = (i_1,i_2,i_3,j_1,j_2,j_3)$. For each subset $T$ of
$\{1,\ldots,6\}$, we define $T$-marginal table $\Bu_{T} =
(u_T(\Bi_T))_{\Bi_T \in {\cal I}_T}$ by
\[
 u_T(\Bi_T) = \sum_{\Bi_{T^{C}} \in {\cal I}_{T^{C}}}u(\Bi_T, \Bi_{T^{C}}), 
\]
where $D^{C}$ denotes the complement of $T$, and ${\cal I}_T$ be the set
of $T$-marginal cells ${\cal I}_T = \prod_{i \in T}\{1,2\}$. Note that
in $u(\Bi_T,\Bi_{T^C})$, the indices in ${\cal I}_T$ are collected to
the left for notation simplicity. These notations are somewhat
unnecessarily exaggerated for our problem, but since they are common in
the literatures of contingency tables, we will use them. The 
notations we use in this section are 
the one-dimensional marginal tables such as 
$\Bu_1 = (u_1(i_1))_{i_1 \in \{1,2\}}$ where
\[
u_1(i_1) =
 \sum_{i_2=1}^2\sum_{i_3=1}^2\sum_{j_1=1}^2\sum_{j_2=1}^2\sum_{j_3=1}^2u(i_1,i_2,i_3,j_1,j_2,j_3),\
 i_1 \in \{1,2\},
\]
the two-dimensional marginal tables such as 
$\Bu_{24} = (u_{24}(i_2,j_1))_{(i_2,j_1)\in \{1,2\}^2}$ where
\[
 u_{24}(i_2,j_1) = 
 \sum_{i_1=1}^2\sum_{i_3=1}^2\sum_{j_1=1}^2\sum_{j_3=1}^2u(i_1,i_2,i_3,j_1,j_2,j_3),\
 (i_2,j_1) \in \{1,2\}^2
\]
and the 
three-dimensional marginal tables such as 
$\Bu_{236} = (u_{236}(i_2,i_3,j_3))_{(i_2,i_3,j_3)\in \{1,2\}^3}$ where
\[
 u_{236}(i_2,i_3,j_3) = 
 \sum_{i_1=1}^2\sum_{j_1=1}^2\sum_{j_2=1}^2u(i_1,i_2,i_3,j_1,j_2,j_3),\
 (i_2,i_3,j_3) \in \{1,2\}^3
\]
and so on.

For each response $\Bu \in \{0,1\}^{64}$ on $D_x^*\times D_y$, we also 
treat $\Bu$ as a response on $D_x\times D_y = \{-1,1\}^6\times
\{-1,1\}^3$ to take the effects of the control factors
$x_4,x_5,x_6$ into consideration.
In that case, we write $\Bu^*$ instead of $\Bu$ and treat
it as a $2^9$ contingency table.
Though somewhat abusing notations, each element of $\Bu^*$ is written as
$u^*(i_1,i_2,i_3,i_4,i_5,i_6,j_1,j_2,j_3)$, and the meaning of 
the subscripts of marginal tables $\Bu_T^*$ changes from $\Bu_T$. For
example, the three-dimensional marginal table $\Bu^*_{236} =
(u_{236}^*(i_2,i_3,i_6))_{(i_2,i_3,i_6) \in \{1,2\}^3}$ means
\[
u_{236}^*(i_2,i_3,i_6) = 
 \sum_{i_1=1}^2\sum_{i_4=1}^2\sum_{i_5=1}^2\sum_{j_1=1}^2\sum_{j_2=1}^2\sum_{j_3=1}^2u^*(i_1,i_2,i_3,i_4,i_5,i_6,j_1,j_2,j_3),\
 (i_2,i_3,i_6) \in \{1,2\}^3. 
\]

Now we consider constraints for the fractions. 
First, the constraints for the size of the designs,
i.e., the constraints meaning that ``for each $(x_1,x_2,x_3) \in
\{-1,1\}^3$, $3$ design points are contained in the fractions'', are
represented as
\begin{equation}
 u_{123}(i_1,i_2,i_3) = 3\ \ \mbox{for}\ \ (i_1,i_2,i_3) \in \{1,2\}^3.
\label{eqn:u123-uniform}
\end{equation}
For late use, we prepare the definition on the uniformity
of the marginal tables.
\begin{definition}
\label{def:uniform}
We define $T$-marginal table $\Bu_T = (u_T(\Bi_T))_{\Bi_T \in {\cal
 I}_T}$ is {\it uniform} if all of $\{u_T(\Bi_T),\ \Bi_T \in {\cal I}_T\}$
 coincides. 
\end{definition} 
Therefore the constraint (\ref{eqn:u123-uniform}) is also written as
``$\Bu_{123}$ is uniform'', under the constraint that the size of the
design is $24$, i.e., $\Bu_{\emptyset} = \Bu^*_{\emptyset} = 24$.   

Now we consider the properties our $24$-runs designs should have. 
Our strategy is to construct $24$-runs designs with uniform marginal
tables that hold for $32$-runs direct product type design 
of Table \ref{tbl:6-3-cross-array-2} as much as possible.
Therefore first we summarize the uniformity of the marginal
tables for the $32$-runs direct product type design.
\begin{proposition}
\label{prop:direct-product}
For the cross-array design with direct product structure given in Table 
\ref{tbl:6-3-cross-array-2}, 
all the two-dimensional marginals of $\Bu^*$ are uniform, and
the
three-dimensional marginals except for $\Bu^*_{124}, \Bu^*_{135},
\Bu^*_{236}, \Bu^*_{456}$ and $\Bu^*_{789}$ are uniform. 
\end{proposition}

\paragraph*{Proof.}\ \ It is easily checked.\hspace*{\fill}$\Box$

\medskip

Proposition \ref{prop:direct-product} means that the design of Table
\ref{tbl:6-3-cross-array-2} is
an orthogonal design of strength $2$, but not of strength $3$. Note that
a design $F$ is {\it orthogonal of strength} $t$, if for any $t$ factors, all
possible combinations of levels appear equally often in $F$. See Chapter
7 of \cite{Wu-Hamada-2009} for detail. 
As we have stated in Section 1, we can estimate 
all the interaction effects between the control factors and the noise
factors for this design, which is 
the principle merit of the direct product type cross-array design. This
important property is guaranteed by the uniformity of
$\Bu^*_{\ell_1\ell_2\ell_3}$ for $1\leq \ell_1 < \ell_2 \leq 6,\ 7\leq
\ell_3 \leq 9$ under the orthogonality of strength $2$. 

Our strategy is to construct designs that have the similar properties 
to Proposition \ref{prop:direct-product} as much as possible 
as $24$-runs designs. Unfortunately, however, it is impossible to
construct $24$-runs designs that can estimate all  
the interaction effects between the control factors and the noise
factors, along with the main effects. This fact is obvious since the
design size $24$ is less than the number of parameters $1 + 9 + 6\times
3 = 28$, i.e., the parameters for interception, main effects, and
interaction effects. For the uniformity of the marginal tables, 
we have the following results.
\begin{proposition}  
\label{prop:uniform-impossible}
For $24$-runs designs of $6$ control factors and $3$ noise factors with
 two-levels,   
\begin{itemize}
\item[(a)] all of 
$\Bu^*_{\ell_1\ell_2\ell_3}$ for $1\leq \ell_1 < \ell_2 \leq 6,\ 7\leq
\ell_3 \leq 9$ cannot be uniform, and 
\item[(b)] all of 
$\Bu^*_{\ell_1\ell_2\ell_3}$ for $1\leq \ell_1 \leq 6,\ 7\leq
\ell_2 < \ell_3 \leq 9$ cannot be uniform.
\end{itemize}
\end{proposition}

\paragraph*{Proof.}\ \ Suppose $\Bu^*_{127}$ and $\Bu^*_{137}$ are
uniform. Then the four-dimensional marginal table $\Bu^*_{1237}$
satisfies
\[
 \sum_{i_3=1}^2u_{1237}(i_1,1,i_3,j_1)
= \sum_{i_3=1}^2u_{1237}(i_1,2,i_3,j_1)
= \sum_{i_2=1}^2u_{1237}(i_1,i_2,1,j_1)
= \sum_{i_2=1}^2u_{1237}(i_1,i_2,2,j_1) = 3
\]
for $(i_1,j_1) \in \{1,2\}^2$. On the other hand, the entries of the
three-dimensional marginal table $\Bu^*_{167}$ is given as
\[
 u^{*}_{167}(i_1,1,j_1) = u^{*}_{1237}(i_1,1,2,j_1) +
 u^{*}_{1237}(i_1,2,1,j_1)
\]
and
\[
 u^{*}_{167}(i_1,2,j_1) = u^{*}_{1237}(i_1,1,1,j_1) +
 u^{*}_{1237}(i_1,2,2,j_1)
\]
for $(i_1,j_1) \in \{1,2\}^2$
from the defining relation $x_6 = x_2x_3$. Therefore we have
\[
 (u_{167}^*(i_1,1,j_1), u_{167}^*(i_1,2,j_1)) \in \{(6,0),\ (4,2),\
 (2,4),\ (0,6)\}, 
\] 
i.e., $\Bu^*_{167}$ cannot be uniform and (a) is proved. (b) is proved
in the same way.\hspace*{\fill}$\Box$

\medskip

From Proposition \ref{prop:uniform-impossible}, we have a choice as to which
three-dimensional marginals are uniform. Of course, 
$\Bu_{124}^*, \Bu_{135}^*, \Bu_{236}^*$ and $\Bu_{456}^*$
cannot be uniform from the defining relation. 
We consider the constraints for the remaining
three-dimensional marginals summarized as follows.

\begin{proposition}
\label{prop:uniform-choice}
$24$-runs designs of $6$ control factors and $3$ noise factors with
two-levels satisfying the following constraints exist.
\begin{itemize}
\item[(a)] All two-dimensional marginal tables of $\Bu^*$ is uniform.
\item[(b)] $\Bu_{124}^*, \Bu_{135}^*, \Bu_{236}^*,
	   \Bu_{456}^*,\Bu_{16\ell}^*, \Bu_{25\ell}^*, \Bu_{34\ell}^*,
	   \Bu_{4\ell_1\ell_2}^*, \Bu_{5\ell_1\ell_2}^*,
	   \Bu_{6\ell_1\ell_2}^*$,\ $\ell = 7,8,9$, 
$7\leq \ell_1 < \ell_2 \leq 9$ are not uniform. All the other
	   three-dimensional marginal tables are uniform.
\end{itemize}
\end{proposition}

\paragraph*{Proof.}\ \ Straightforward from Proposition
	   \ref{prop:uniform-impossible}. \hspace*{\fill}$\Box$

\medskip

The reason to consider the constraints of this type 
is revealed below. 
We can show that the designs satisfying the conditions of
Proposition \ref{prop:uniform-choice} can be characterized as the
orthogonal fractions of strength $3$ for $x_1,x_2,x_3,y_1,y_2,y_3$. 

\begin{theorem}
\label{thm:equiv-st3-6factors}
$24$-runs designs of $6$ control factors and $3$ noise factors with
two-levels satisfy the constraints of 
Proposition \ref{prop:uniform-choice} if and only if all
 three-dimensional marginal tables of $\Bu$ are uniform.
\end{theorem} 

\paragraph*{Proof.}\ \ By calculating the summations, it is
straightforward to show that $\Bu^*$
satisfies the constraints of 
Proposition \ref{prop:uniform-choice} if all three-dimensional marginal
tables of $\Bu$ are uniform. To show the reverse direction, 
we show that the uniformity of $\Bu_{267}^*$ can
be removed from the constraints of 
Proposition \ref{prop:uniform-choice}. From the uniformity of
$\Bu^*_{237}$,  
the four-dimensional marginal tables $\Bu^*_{2367}$ satisfies
\[
\sum_{i_6=1}^2 u_{2367}^*(i_2,i_3,i_6,i_7) = 3,\ \ (i_2,i_3,i_7) \in \{1,2\}^3.
\]
From the defining relation $x_2x_3 = x_6$, 
the four-dimensional marginal tables $\Bu^*_{2367}$ also satisfies
\[\begin{array}{c}
\displaystyle\sum_{i_7=1}^2 u_{2367}^*(i_2,i_3,1,i_7) = 6,\ \
\displaystyle\sum_{i_7=1}^2 u_{2367}^*(i_2,i_3,2,i_7) = 0,\ \
 (i_2,i_3) \in \{(1,2),(2,1)\},\\
\displaystyle\sum_{i_7=1}^2 u_{2367}^*(i_2,i_3,1,i_7) = 0,\ \
\displaystyle\sum_{i_7=1}^2 u_{2367}^*(i_2,i_3,2,i_7) = 6,\ \
 (i_2,i_3) \in \{(1,1),(2,2)\}.
\end{array}
\]
These relation determines $\Bu^*_{2367}$ uniquely as
\[\begin{array}{c}
\displaystyle 
 u_{2367}^*(i_2,i_3,1,i_7) = 3,\ \ 
 u_{2367}^*(i_2,i_3,2,i_7) = 0,\ \ 
(i_2,i_3) \in \{(1,2),(2,1)\}, i_7=1,2,\\
\displaystyle 
 u_{2367}^*(i_2,i_3,1,i_7) = 0,\ \ 
 u_{2367}^*(i_2,i_3,2,i_7) = 3,\ \ 
(i_2,i_3) \in \{(1,1),(2,2)\}, i_7=1,2,\\
\end{array}
\]
which yields that $\Bu_{267}^*$ is uniform. All the constraints including
$x_4,x_5,x_6$, i.e., $\Bu_{14\ell}^*$, $\Bu_{15\ell}^*$, $\Bu_{24\ell}^*$,
$\Bu_{26\ell}^*$, $\Bu_{35\ell}^*$, $\Bu_{36\ell}^*$, $\Bu_{45\ell}^*$,
$\Bu_{46\ell}^*$ and $\Bu_{56\ell}^*$ for $\ell = 7,8,9$,  
can be removed similarly.
\hspace*{\fill}$\Box$

\medskip

Now we can define the ideal to enumerate designs as the coefficients of
the corresponding indicator functions. From Theorem
\ref{thm:equiv-st3-6factors}, it is sufficient to consider the algebraic
equations for $2^6=64$ variables, not for $2^9 = 512$ variables.
Because our problem is for two-level factors, 
the algebraic equations to solve
is given by 
(\ref{eqn:2-level-equations}) 
with constraints. The constraints of Theorem
\ref{thm:equiv-st3-6factors} 
is written in the form of $C\Bu = CX\Btheta = \Bh$. For
example, the constraint that $\Bu_{123}$ is uniform, i.e., 
$u_{123}(i_1,i_2,i_3) = 3$ for $(i_1,i_2,i_3) \in \{1,2\}^3$, 
is written in the form of $CX\Btheta = \Bh$ for $C = E_8
\otimes \Bone_8^T$ and $\Bh = 3\Bone_8$, where $E_8$ denote an $8\times
8$ identity matrix,
$\Bone_8 = (1,1,1,1,1,1,1,1)^T$ and $\otimes$ denote a Kronecker product.

\begin{remark}
We express the constraints as $CX\Btheta = \Bh$ with the general cases
 in mind. Since our problem is the two-level case, the structure of the
 indicator functions for two-level cases given by 
\cite{Fontana-Pistone-Rogantin-2000} can be used. In fact, the
 constraints of Theorem \ref{thm:equiv-st3-6factors} is simply written as
\begin{equation}
 \theta_{000000} = \frac{3}{8},\ \theta_{a_1a_2a_3b_1b_2b_3} = 0\  \
 \mbox{for}\ \  a_1+a_2+a_3+b_1+b_2+b_3 = 1,2,3.
\label{eqn:const-our-problems}
\end{equation}
In addition, 
by substituting
(\ref{eqn:const-our-problems}) to (\ref{eqn:2-level-equations}),  we
also obtain 
\[
\theta_{a_1a_2a_3b_1b_2b_3} = 0\  \
 \mbox{for}\ \  a_1+a_2+a_3+b_1+b_2+b_3 = 5,6. 
\]
We can also check these results from the output of the algebraic computation.
\end{remark}

\begin{remark}
From an application point of view, the constraint of the uniformity for
 $\Bu_{456}$ is not
 so important, since the interaction effects between the noise factors
 are usually ignored. However, by actual computation, we find that all
 the solutions without the
 uniformity of $\Bu_{456}$ also satisfy the uniformity of $\Bu_{456}$. 
\end{remark}

The calculations are done by Singular (\cite{Singular})
installed in MacBook Pro, 2.3 GHz, Quad-Core, Intel Core i7. 
We use the command \verb|minAssGTZ| to compute minimal primes of the ideal.
After 
calculations within one minute, we find $192$ solutions. We also find
that all the solutions are classified into the same equivalence class
for sign changes of levels and permutations of factors within control
factors and within noise factors. One of the solutions is already shown
in Table \ref{tbl:6-3-new-design}, with the indicator function
\[
 \begin{array}{rl}
\displaystyle 
f(x_1,x_2,x_3,y_1,y_2,y_3) = & 
\displaystyle 
\frac{3}{8} + \frac{1}{8}\left(
x_1x_2x_3y_1 + x_1x_2x_3y_2 - x_1x_2x_3y_3 +x_1x_2y_1y_2 + x_1x_2y_1y_3\right.\\
& {}- x_1x_2y_2y_3+x_1x_3y_1y_2 + x_1x_3y_1y_3+x_1x_3y_2y_3+x_2x_3y_1y_2\\
&
 \left.{} - x_2x_3y_1y_3 + x_2x_3y_2y_3 - x_1y_1y_2y_3 + x_2y_1y_2y_3 -
  x_3y_1y_2y_3\right).
\end{array}
\]
Although redundant, we also show some other solutions. By permuting
noise factors and their levels, all the solution can be converted so
that two points 
$(x_1,x_2,x_3,y_1,y_2,y_3)
 = (-1,-1,-1,-1,-1,-1)$ and $(-1,-1,-1,-1,-1,1)$ are included in the
 design. We find there are $12$ such solutions of Table 
\ref{tbl:12-designs-solution}, where $F_1$ is the design shown in Table 
\ref{tbl:6-3-new-design}.
\begin{table}[htbp]
\caption{The list of the indicator functions of $24$-runs designs including
 $(x_1,x_2,x_3,y_1,y_2,y_3)
 = (-1,-1,-1,-1,-1,-1)$ and $(-1,-1,-1,-1,-1,1)$. (All the other
 coefficients are zero. $F_1$ is displayed in Table
\ref{tbl:6-3-new-design}.)}
\label{tbl:12-designs-solution}
\[
 \begin{array}{|c|r|r|r|r|r|r|r|r|r|r|r|r|}\hline
& F_1 & F_2 & F_3 & F_4 & F_5 & F_6 & F_7 & F_8 & F_9 & F_{10} & F_{11} &
 F_{12}\\ \hline
\theta_{000000} & 3/8& 3/8& 3/8& 3/8& 3/8& 3/8& 3/8& 3/8& 3/8& 3/8& 3/8& 3/8\\
\theta_{111100} & 1/8& 1/8& 1/8& 1/8& 1/8& 1/8& 1/8& 1/8& 1/8& 1/8& 1/8& 1/8\\
\theta_{111010} & 1/8& 1/8& 1/8& 1/8& 1/8& 1/8& 1/8& 1/8& 1/8& 1/8& 1/8& 1/8\\
\theta_{111001} &-1/8& 1/8&-1/8& 1/8&-1/8& 1/8&-1/8& 1/8&-1/8&-1/8& 1/8& 1/8\\
\theta_{110110} & 1/8& 1/8& 1/8& 1/8& 1/8& 1/8& 1/8& 1/8& 1/8& 1/8& 1/8& 1/8\\
\theta_{110101} & 1/8&-1/8&-1/8&-1/8& 1/8&-1/8& 1/8& 1/8&-1/8& 1/8&-1/8& 1/8\\
\theta_{110011} &-1/8&-1/8& 1/8&-1/8& 1/8& 1/8& 1/8&-1/8& 1/8&-1/8& 1/8&-1/8\\
\theta_{101110} & 1/8& 1/8& 1/8& 1/8& 1/8& 1/8& 1/8& 1/8& 1/8& 1/8& 1/8& 1/8\\
\theta_{101101} & 1/8&-1/8& 1/8& 1/8&-1/8&-1/8& 1/8&-1/8& 1/8&-1/8& 1/8&-1/8\\
\theta_{101011} & 1/8& 1/8& 1/8&-1/8& 1/8&-1/8&-1/8&-1/8&-1/8& 1/8&-1/8& 1/8\\
\theta_{011110} & 1/8& 1/8& 1/8& 1/8& 1/8& 1/8& 1/8& 1/8& 1/8& 1/8& 1/8& 1/8\\
\theta_{011101} &-1/8& 1/8& 1/8&-1/8& 1/8& 1/8&-1/8&-1/8& 1/8& 1/8&-1/8&-1/8\\
\theta_{011011} & 1/8&-1/8&-1/8& 1/8&-1/8&-1/8& 1/8& 1/8& 1/8& 1/8&-1/8&-1/8\\
\theta_{100111} &-1/8& 1/8&-1/8& 1/8&-1/8& 1/8&-1/8& 1/8& 1/8& 1/8&-1/8&-1/8\\
\theta_{010111} & 1/8& 1/8& 1/8& 1/8&-1/8&-1/8&-1/8&-1/8&-1/8&-1/8& 1/8& 1/8\\
\theta_{001111} &-1/8&-1/8&-1/8&-1/8& 1/8& 1/8& 1/8& 1/8&-1/8&-1/8& 1/8& 1/8\\ \hline
 \end{array}
\]
\end{table}

Finally in this paper, we consider estimable parameters under the
$24$-runs design $F_1$. By the calculation of a reduced Gr\"obner basis of the
design ideal $I(F_1)$, we find
a basis of
$\mathbb{Q}[x_1,x_2,x_3,x_4,x_5,x_6,y_1,y_2,y_3]/I(F_1)$ as follows.
\[\begin{array}{cccccccc}
1, & x_1, & x_2, & x_3, & x_4,& x_5,& x_6, & y_1,\\
y_2, & y_3, & x_1y_2, & x_1y_3, & x_2y_2, & x_2y_3, & x_3y_3, & x_4y_3,\\
x_5y_1, & x_5y_3, & x_5y_2, & x_6y_2, & x_6y_3, & y_1y_2, & y_1y_3, & y_2y_3
\end{array}
\] 
This basis is derived from the reduced Gr\"obner basis under the reverse
lexicographic term order. Since our interest is to estimate interaction
effect between the control factors and the noise factors, adding to the
main effects for these factors, we investigate the statistical model of
the form
\begin{equation}
 \mu + \sum_{i = i}^6\alpha_ix_i + \sum_{j = 1}^3\beta_jy_j + \sum_{i =
 1}^6\sum_{j = 1}^3\gamma_{ij}x_iy_j,
\label{eqn:statistical-model}
\end{equation}
where $\mu$ be the interception, 
$\alpha_i$ be the main effect of $x_i$, $\beta_j$ be the main
effect of $y_j$, and $\gamma_{ij}$ be the interaction effect between
$x_i$ and $y_j$. Dividing (\ref{eqn:statistical-model}) by the reduced
Gr\"obner basis of $I(F_1)$, we find the following standard form.
\begin{equation}
 \begin{array}{l}
\mu + \alpha_1x_1 + \alpha_2x_2 + \alpha_3x_3 + (\alpha_4 + \gamma_{32})x_4
+ (\alpha_5 + \gamma_{11} + \gamma_{21} - \gamma_{31} + \gamma_{32})x_5
+ (\alpha_6 + \gamma_{11})x_6\\
\hspace*{5mm}{}+ (\beta_1 - \gamma_{61})y_1
+ (\beta_2 - \gamma_{42})y_2
+ (\beta_3 - \gamma_{42} - \gamma_{61})y_3 + (\gamma_{12} +
\gamma_{21})x_1y_2 + (\gamma_{13} + \gamma_{31})x_1y_3\\
\hspace*{5mm}{}+ (\gamma_{22} + \gamma_{31})x_2y_2
+ (\gamma_{11} + \gamma_{21} + \gamma_{23} - \gamma_{31} + \gamma_{32})x_2y_3
+ (-\gamma_{21} + \gamma_{33})x_3y_3\\
\hspace*{5mm}{}+ (-\gamma_{41} - \gamma_{42} + \gamma_{43})x_4y_3
+ (-\gamma_{41} + \gamma_{51})x_5y_1
+ (\gamma_{41} + \gamma_{42} + \gamma_{52} + \gamma_{61})x_5y_2\\
\hspace*{5mm}{}+ (\gamma_{42} + \gamma_{53} + \gamma_{61})x_5y_3
+ (\gamma_{41} + \gamma_{42} + \gamma_{61} + \gamma_{62})x_6y_2
+ (\gamma_{41} + \gamma_{42} + \gamma_{63})x_6y_3\\
\hspace*{5mm}{}
+ (-\gamma_{11} - \gamma_{21} + \gamma_{31} - \gamma_{32})y_1y_2
-\gamma_{32}y_1y_3-\gamma_{11}y_2y_3
 \end{array}
\label{eqn:confounding-relations-F2}
\end{equation}
This standard form shows the confounding relations of the parameters in
(\ref{eqn:statistical-model}) under $F_1$. To see the estimability of
the parameters, we make a matrix from the relation 
(\ref{eqn:confounding-relations-F2}) as follows.
\[{\footnotesize
\left[
 \begin{array}{r|rrrrrr|rrr|rrr|rrr|rrr|rrr|rrr|rrr}
1& 0& 0& 0& 0& 0& 0& 0& 0& 0& 0& 0& 0& 0& 0& 0& 0& 0& 0& 0& 0& 0& 0& 0&
 0& 0& 0& 0\\
0& 1& 0& 0& 0& 0& 0& 0& 0& 0& 0& 0& 0& 0& 0& 0& 0& 0& 0& 0& 0& 0& 0& 0&
 0& 0& 0& 0\\
0& 0& 1& 0& 0& 0& 0& 0& 0& 0& 0& 0& 0& 0& 0& 0& 0& 0& 0& 0& 0& 0& 0& 0&
 0& 0& 0& 0\\
0& 0& 0& 1& 0& 0& 0& 0& 0& 0& 0& 0& 0& 0& 0& 0& 0& 0& 0& 0& 0& 0& 0& 0&
 0& 0& 0& 0\\
0& 0& 0& 0& 1& 0& 0& 0& 0& 0& 0& 0& 0& 0& 0& 0& 0& 1& 0& 0& 0& 0& 0& 0&
 0& 0& 0& 0\\
0& 0& 0& 0& 0& 1& 0& 0& 0& 0& 1& 0& 0& 1& 0& 0&-1& 1& 0& 0& 0& 0& 0& 0&
 0& 0& 0& 0\\
0& 0& 0& 0& 0& 0& 1& 0& 0& 0& 1& 0& 0& 0& 0& 0& 0& 0& 0& 0& 0& 0& 0& 0&
 0& 0& 0& 0\\
0& 0& 0& 0& 0& 0& 0& 1& 0& 0& 0& 0& 0& 0& 0& 0& 0& 0& 0& 0& 0& 0& 0& 0&
 0&-1& 0& 0\\
0& 0& 0& 0& 0& 0& 0& 0& 1& 0& 0& 0& 0& 0& 0& 0& 0& 0& 0& 0&-1& 0& 0& 0&
 0& 0& 0& 0\\
0& 0& 0& 0& 0& 0& 0& 0& 0& 1& 0& 0& 0& 0& 0& 0& 0& 0& 0& 0&-1& 0& 0& 0&
 0&-1& 0& 0\\
0& 0& 0& 0& 0& 0& 0& 0& 0& 0& 0& 1& 0& 1& 0& 0& 0& 0& 0& 0& 0& 0& 0& 0&
 0& 0& 0& 0\\
0& 0& 0& 0& 0& 0& 0& 0& 0& 0& 0& 0& 1& 0& 0& 0& 1& 0& 0& 0& 0& 0& 0& 0&
 0& 0& 0& 0\\
0& 0& 0& 0& 0& 0& 0& 0& 0& 0& 0& 0& 0& 0& 1& 0& 1& 0& 0& 0& 0& 0& 0& 0&
 0& 0& 0& 0\\
0& 0& 0& 0& 0& 0& 0& 0& 0& 0& 1& 0& 0& 1& 0& 1&-1& 1& 0& 0& 0& 0& 0& 0&
 0& 0& 0& 0\\
0& 0& 0& 0& 0& 0& 0& 0& 0& 0& 0& 0& 0&-1& 0& 0& 0& 0& 1& 0& 0& 0& 0& 0&
 0& 0& 0& 0\\
0& 0& 0& 0& 0& 0& 0& 0& 0& 0& 0& 0& 0& 0& 0& 0& 0& 0& 0&-1&-1& 1& 0& 0&
 0& 0& 0& 0\\
0& 0& 0& 0& 0& 0& 0& 0& 0& 0& 0& 0& 0& 0& 0& 0& 0& 0& 0&-1& 0& 0& 1& 0&
 0& 0& 0& 0\\
0& 0& 0& 0& 0& 0& 0& 0& 0& 0& 0& 0& 0& 0& 0& 0& 0& 0& 0& 1& 1& 0& 0& 1&
 0& 1& 0& 0\\
0& 0& 0& 0& 0& 0& 0& 0& 0& 0& 0& 0& 0& 0& 0& 0& 0& 0& 0& 0& 1& 0& 0& 0&
 1& 1& 0& 0\\
0& 0& 0& 0& 0& 0& 0& 0& 0& 0& 0& 0& 0& 0& 0& 0& 0& 0& 0& 1& 1& 0& 0& 0&
 0& 1& 1& 0\\
0& 0& 0& 0& 0& 0& 0& 0& 0& 0& 0& 0& 0& 0& 0& 0& 0& 0& 0& 1& 1& 0& 0& 0&
 0& 0& 0& 1\\
0& 0& 0& 0& 0& 0& 0& 0& 0& 0&-1& 0& 0&-1& 0& 0& 1&-1& 0& 0& 0& 0& 0& 0&
 0& 0& 0& 0\\
0& 0& 0& 0& 0& 0& 0& 0& 0& 0& 0& 0& 0& 0& 0& 0& 0&-1& 0& 0& 0& 0& 0& 0&
 0& 0& 0& 0\\
0& 0& 0& 0& 0& 0& 0& 0& 0& 0&-1& 0& 0& 0& 0& 0& 0& 0& 0& 0& 0& 0& 0& 0&
 0& 0& 0& 0
 \end{array}
\right]
}\]
Note that each column corresponds to each parameter, and each row
corresponds to each term of (\ref{eqn:confounding-relations-F2}). 
Then the statistical model with estimable parameters is constructed by
choosing columns from the above matrix so that they are linearly
independent. See Chapter 3.7 of \cite{Pistone-Riccomagno-Wynn-2001} for detail. 
For example, we see that the set of the $24$ columns
obtained by removing $19,25,27$ and $28$-th columns from the above matrix is
linearly independent. Therefore the parameters in the corresponding 
statistical model
\begin{equation}
\begin{array}{l}
\displaystyle
 \mu + \sum_{i = i}^6\alpha_ix_i + \sum_{j = 1}^3\beta_jy_j 
+ \gamma_{11}x_1y_1 + \gamma_{12}x_1y_2 + \gamma_{13}x_1y_3
+ \gamma_{21}x_2y_1 + \gamma_{22}x_2y_2 + \gamma_{23}x_2y_3\\
\hspace*{5mm}{}+ \gamma_{31}x_3y_1 + \gamma_{32}x_3y_2 
+ \gamma_{41}x_4y_1 + \gamma_{42}x_4y_2 + \gamma_{43}x_4y_3
+ \gamma_{51}x_5y_1 + \gamma_{52}x_5y_2 
+ \gamma_{61}x_6y_1 
\end{array}
\label{eqn:saturated-model-example}
\end{equation}
are estimable.

\section{Discussion}
In this paper, we consider using the indicator functions to
enumerate cross-array designs with some desirable properties. Although
the theory of the indicator functions to enumerate designs is already given
in previous papers, the computational feasibility becomes a problem in
practice. In this paper, to avoid calculation for $2^9 = 512$ variables
in a naive approach, we define the ideal for $2^6 = 64$ variables by
considering constraints to have in the designs. 
It is true that our result is not general, i.e., we consider 
only one setting,
$24$-runs
designs for $6$ control factors and $3$ noise factor. However, 
we think that our approach can be applicable to other settings.
As another contribution, the $24$-runs design obtained in this paper
is valuable in application. By the obtained $24$-runs design, we can
estimate all the parameters in the statistical model with (up to) $14$
interaction effects and all the main effects such as
(\ref{eqn:saturated-model-example}). 

Somewhat interestingly, 
looking at Table \ref{tbl:6-3-new-design}, we find the ``point symmetry''
for the symbol $\circ$. It is easy to check that the group action for
level changes and permutations of factors preserve this point
symmetry. Therefore, all the $192$ solutions have this point symmetry.

\bibliographystyle{plain}

\end{document}